\documentclass[aps,pra,reprint,amsmath,amssymb,citeautoscript,superscriptaddress]{revtex4-1}

\usepackage{amsmath} 
\usepackage{graphicx} 
\usepackage{amssymb}
\usepackage{amsthm}
\usepackage{bm}
\usepackage[usenames,dvipsnames]{xcolor}
\definecolor{myblue}{rgb}{0,0,1}
\usepackage[colorlinks=true,linkcolor=myblue,urlcolor=myblue,citecolor=myblue]{hyperref}

\usepackage[T1]{fontenc}
\usepackage{txfonts}

\let\vr\undefined

\newcommand{\vr}{{\bm{r}}}
\newcommand{\vq}{{\bm{q}}}
\newcommand{\vp}{{\bm{p}}}
\newcommand{\vP}{{\bm{P}}}
\newcommand{\vk}{{\bm{k}}}

\begin{document}

\title{
Bright and dark singlet excitons via linear and two-photon spectroscopy
in monolayer transition-metal dichalcogenides
}

\author{Timothy C. Berkelbach}
\affiliation{Princeton Center for Theoretical Science, Princeton University, 
Princeton, NJ 08544, USA}

\author{Mark S. Hybertsen}
\affiliation{Center for Functional Nanomaterials, Brookhaven National Laboratory, 
Upton, NY 11973-5000, USA}

\author{David R. Reichman}
\affiliation{Department of Chemistry, Columbia University, New York, NY 10027, USA}

\begin{abstract}
We discuss the linear and two-photon spectroscopic selection rules for
spin-singlet excitons in monolayer transition-metal dichalcogenides.  Our
microscopic formalism combines a fully $k$-dependent few-orbital band structure
with a many-body Bethe-Salpeter equation treatment of the electron-hole
interaction, using a model dielectric function.  We show analytically and
numerically that the single-particle, valley-dependent selection rules are
preserved in the presence of excitonic effects. Furthermore, we definitively
demonstrate that the bright (one-photon allowed) excitons have $s$-type
azimuthal symmetry and that dark $p$-type excitons can be probed via two-photon
spectroscopy.  The screened Coulomb interaction in these materials
substantially deviates from the $1/\varepsilon_0 r$ form; this breaks the
``accidental'' angular momentum degeneracy in the exciton spectrum, such that
the 2$p$ exciton has a lower energy than the 2$s$ exciton by at least 50 meV.
We compare our calculated two-photon absorption spectra to recent experimental
measurements.
\end{abstract}

\maketitle

\section{Introduction}

The transition-metal dichalcogenides (TMDCs) are a family of layered
semiconducting crystals that includes MoS$_2$, MoSe$_2$, WS$_2$, and WSe$_2$.
Isolated monolayers of TMDCs have been recently investigated for two major
reasons.  First, the emergent direct band-gap occurs at the corners of the
hexagonal Brillouin zone (so-called `valleys')~\cite{spl10,mak10} and the nearby
band structure topology leads to valley-dependent optical selection
rules~\cite{xia12,zen12,mak12}.  Second, the carrier confinement and reduced
dielectric screening leads to large many-body effects, such as the formation of
strongly bound excitons~\cite{ber13,qiu13,berg14,he14,che14},
trions~\cite{mak13,ros13,jon13,ber13}, and biexcitons~\cite{you15} with very
large binding energies.  A unified understanding of the optical properties must
treat both of these aspects on equal footing, and significant effort is now
being focused on investigating the detailed spectroscopy of excitons in
monolayer TMDCs.

In the ongoing effort to understand excitons in these materials, multiple
spectroscopic techniques have been employed, including
reflectance~\cite{he14,che14,li14}, photoluminescence excitation
spectroscopy~\cite{hil15} scanning tunneling
spectroscopy~\cite{zha14sts,uge14}, and two-photon
luminescence~\cite{he14,ye14,wan15}.  A rigorous knowledge of the spectroscopic
selection rules for excitons in monolayer TMDCs is crucial for the proper
interpretation of these and future experiments.  In this paper, we develop a
model-based framework which is sufficiently detailed to provide quantitative
results, but also sufficiently simple to allow precise statements about
symmetry-determined selection rules.  We describe the connection to our
previous work based on an effective mass theory of excitons~\cite{ber13}, and
identify the key microscopic physical factors that determine the properties of
excitons and their interaction with photons.  We also provide the first
theoretical treatment of two-photon absorption in monolayer TMDCs.

The outline of the paper is as follows.  In Sec.~\ref{sec:band} we will discuss
simple microscopic models of the single-particle band structure in monolayer
TMDCs, and in particular we will analyze the transition matrix elements which
\textit{completely} determine the independent-electron absorption and
\textit{partially} determine the excitonic absorption.  We will then in
Sec.~\ref{sec:linear} analyze the linear optical properties and present
selection rules, both in the absence and presence of exciton effects,
definitively finding that $s$-type excitons are optically bright.  Lastly, in
Sec.~\ref{sec:tpa} we will calculate the two-photon absorption signal which
will be shown to probe $p$-type excitons and we will discuss some of the
implications for recent experiments.  We conclude in Sec.~\ref{sec:conc}, and
make connection to other recent theoretical works.  We note that a preliminary
version of this work appeared in Ref.~\cite{ber14thesis}.

\section{Single-particle band structure}
\label{sec:band}

We will consider two models for the single-particle band structure.  First, we
will consider a widely used long-wavelength, two-band model~\cite{xia12}.  In
particular, this minimal model allows for a largely analytical treatment, which
exposes many of the subtleties of the theory, including selection rules and
exciton effects.  Second, we will use a recently presented nonlinear three-band
model~\cite{liu13}, which requires a numerical treatment but captures
higher-order effects.  This also ensures that our conclusions are generally
valid and not specifically dependent on the simplified two-band picture.  For
simplicity we will henceforth neglect spin-orbit coupling, though it can be
straightforwardly included in the single-particle
description~\cite{xia12,kor13,liu13}.  Specifically, in all models of the band
structure, the spin projection $s_z$ is still a good quantum number in the
presence of spin-orbit coupling. In this sense, the following discussion
applies to the $A$-exciton (and not the $B$-exciton) and conventional factors
of two for spin will not appear.  At this level of theory, the formalism for
the $B$-exciton is identical, and its contribution is simply shifted to higher
energies.

\subsection{Two-band model}

The first model considered has the form of a conventional two-band, massive Dirac Hamiltonian,
\begin{equation}
H_\tau(\vk) = \left(
\begin{array}{cc}
E_g/2           & at(\tau q_x - iq_y) \\
at(\tau q_x+iq_y)  & -E_g/2
\end{array}
\right).
\end{equation}
The variable $\tau=\pm 1$ indexes the two ``valleys,'' known as the $K$ and
$K^\prime$ (or $K$ and $-K$) points, which occur at alternating corners of the
hexagonal first Brillouin zone. The Hamiltonian has been linearized in the
wavevector difference with respect to the nearest $K$ point, i.e.
$\vq=\vk-\bm{K}$.  This is a gapped version of the conventional graphene
Hamiltonian~\cite{cas09}.  In graphene, the spinor basis corresponds to carbon
$p_z$ orbitals on the two distinct sublattices; in the TMDCs, the basis
corresponds to the transition-metal $|d_{z^2}\rangle \equiv |\phi_c\rangle$
orbital and the metal symmetry-adapted $|d_{x^2-y^2}\rangle+i\tau|d_{xy}\rangle
\equiv |\phi_v^{\tau}\rangle$ orbital.  The above Hamiltonian was first used
for TMDCs by Xiao et al.~\cite{xia12} who predicted optical selection rules
leading to spin-valley coupling. Such spin-valley coupling was quickly
confirmed experimentally, by monitoring the circular polarization of
photoluminescence~\cite{zen12,mak12}. 

The eigenvalues of the two-band Hamiltonian are
\begin{equation}
E_{c/v}(\vk) = \pm \frac{1}{2} \sqrt{E_g^2 + 4 (atq)^2} \equiv \pm \varepsilon(\vk)
\end{equation}
and the eigenvectors are
\begin{subequations}
\begin{align}
|\psi_{c,\vk}^\tau\rangle &= \frac{1}{\sqrt{2}}\left[ \hspace{0.7em} 
    \sqrt{\alpha_+}(\vk) |\phi_c\rangle 
    + \sqrt{\alpha_-}(\vk) e^{i\tau\phi_\vk} |\phi_v^\tau\rangle \right] \\
|\psi_{v,\vk}^\tau\rangle &= \frac{1}{\sqrt{2}}\left[
    - \sqrt{\alpha_-}(\vk) |\phi_c\rangle 
    + \sqrt{\alpha_+}(\vk) e^{i\tau\phi_\vk} |\phi_v^\tau\rangle \right].
\end{align}
\end{subequations}
where $\alpha_{\pm}(\vk) = 1\pm E_g/[2\varepsilon(\vk)]$ and $\tan \phi_\vk =
q_y/q_x$.  The \textit{relative} phase appearing within each eigenvector is
associated with an electronic ``chirality'' (related to Berry's phase), which
is well-known in graphene~\cite{cas09,gei07,bar07}.  Note that the
\textit{overall} phase of each eigenvector is arbitrary, and the phase
convention chosen here is such that the first element of each eigenvector is
purely real.

\subsection{Three-band model}

\begin{figure}[t!]
\centering
\includegraphics[scale=0.95]{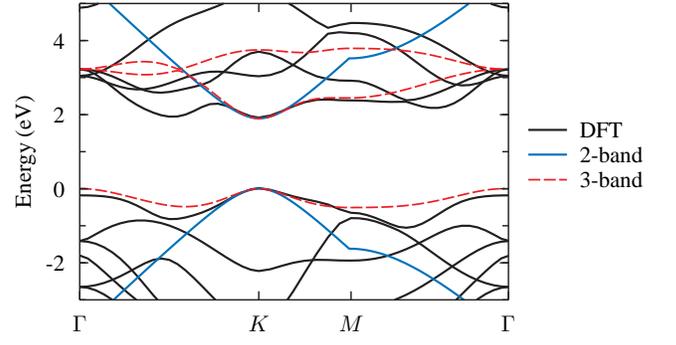}
\caption{Single-particle band structure of MoS$_2$ predicted by a linearized
two-band model (blue solid) and a non-linear three-band model (red dashed)
compared to first-principles density functional theory with the local density
approximation (DFT, solid black).
}
\label{fig:bands}
\end{figure}

A more detailed Hamiltonian -- using three bands derived from the
transition-metal $|d_{z^2}\rangle$, $|d_{xy}\rangle$, and $|d_{x^2-y^2}\rangle$
atomic orbitals -- was given recently by Liu et al~\cite{liu13}.  The form of
the matrix elements and material-specific parameters can be found in
Ref.~\cite{liu13}. We note than in addition to using three bands instead
of two, this Hamiltonian has \textit{not} been linearized with respect to
wavevector near the $K$ and $K^\prime$ points, which gives a more accurate
description throughout the entire Brillouin zone.  While it cannot be so easily
diagonalized analytically, the Hamiltonian can be straightforwardly
diagonalized numerically. For phase consistency in later calculations, we
enforce the same phase convention as for the two-band eigenvectors, i.e.~that
the first element of each eigenvector is purely real, which is sufficient to
ensure continuity in $k$-space.  In Fig.~\ref{fig:bands}, the band structure
predicted by these two models is compared to the band structure calculated by
density functional theory with the local density approximation.

\begin{figure*}
\centering
\includegraphics[scale=0.35]{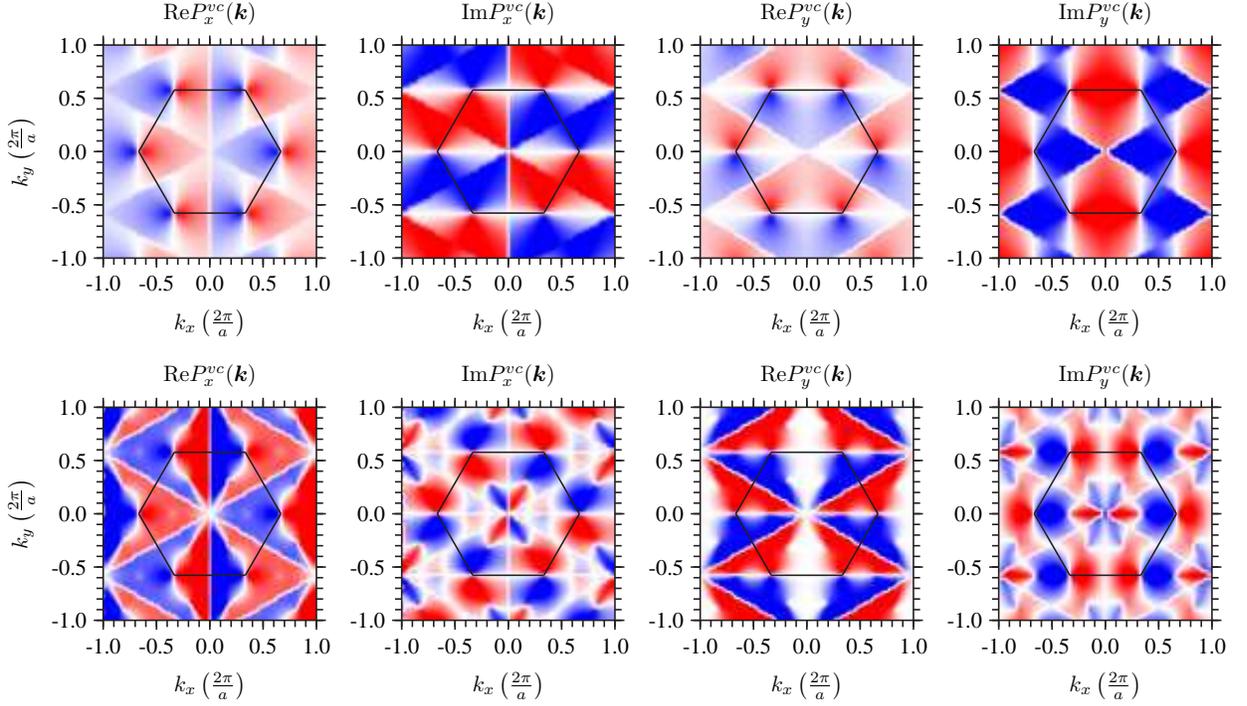}
\caption{Valence to lowest conduction band momentum matrix elements for a linearized
two-band model (top) and a non-linear three-band model (bottom).  Blue is
positive, red is negative, and white is zero.  The results are qualitatively
very similar in the immediate vicinity of the $K$ and $K^\prime$ points, but
differ elsewhere in the Brillouin zone.
}
\label{fig:pcv_k}
\end{figure*}

\subsection{Transition matrix elements}
\label{ssec:trans}

An analysis of optical selection rules requires the momentum matrix elements
between single-particle states.  In the present model Hamiltonians, the
momentum matrix elements normal to the layer are zero by symmetry.  Here we
focus on the momentum in the plane.  By using the commutation relation $\vp =
(-im/\hbar) [\vr,H]$, we can write these momentum matrix elements as
\begin{equation}
\begin{split}
\vP^{vc}(\vk) &= \frac{-im}{\hbar}
    \langle \psi_{v,\vk} | \left[\vr,H\right] | \psi_{c,\vk} \rangle \\
    &= \frac{m}{\hbar} \left(E_{c,\vk}-E_{v,\vk}\right) 
        \langle \psi_{v,\vk} | \nabla_\vk | \psi_{c,\vk} \rangle
\end{split}
\end{equation}
where we have used the $k$-space representation of the position operator, $\vr
= i\nabla_\vk$. We can now use a generalized Feynman-Hellman theorem to write
this as
\begin{equation}
\vP^{vc}(\vk) = \frac{m}{\hbar} \langle \psi_{v,\vk} | \nabla_\vk H(\vk) | \psi_{c,\vk} \rangle
\end{equation}
(note that this expression neglects the on-site, intra-atomic
contribution~\cite{ped01}, however this vanishes here for $d-d$ transitions).
For the simple two-band Hamiltonian, this gives
\begin{equation}
\nabla_\vk H(\vk) = \left(
\begin{array}{cc}
0                          & at(\tau \hat{x} - i\hat{y}) \\
at(\tau \hat{x}+i\hat{y})  & 0
\end{array}
\right).
\end{equation}
The appropriate matrix element can then be taken between the conduction and
valence band eigenstates of the Hamiltonian, yielding a transition dipole
vector $\vP^{vc}(\vk)$ with linear $x$- and $y$-polarization components
\begin{align}
P^{vc}_{x}(\vk) &= \tau \frac{mat}{2\hbar} 
    \left[\alpha_+(\vk)e^{-i\tau\phi_\vk} 
        - \alpha_-(\vk)e^{i\tau\phi_\vk} \right], \\
P^{vc}_{y}(\vk) &= \hspace{0.2em} i \frac{mat}{2\hbar} 
    \left[\alpha_+(\vk)e^{-i\tau\phi_\vk} 
        + \alpha_-(\vk)e^{i\tau\phi_\vk} \right].
\end{align}
The same procedure can be done for the three-band Hamiltonian, by taking the
gradient and calculating (numerically) the appropriate matrix element between
conduction and valence bands.  A comparison of the real and imaginary parts of
the $x$- and $y$-components of the two different models of the band structure
is shown in Fig.~\ref{fig:pcv_k} throughout the entire first Brillouin zone.  

\begin{figure}
\centering
\includegraphics[scale=0.36]{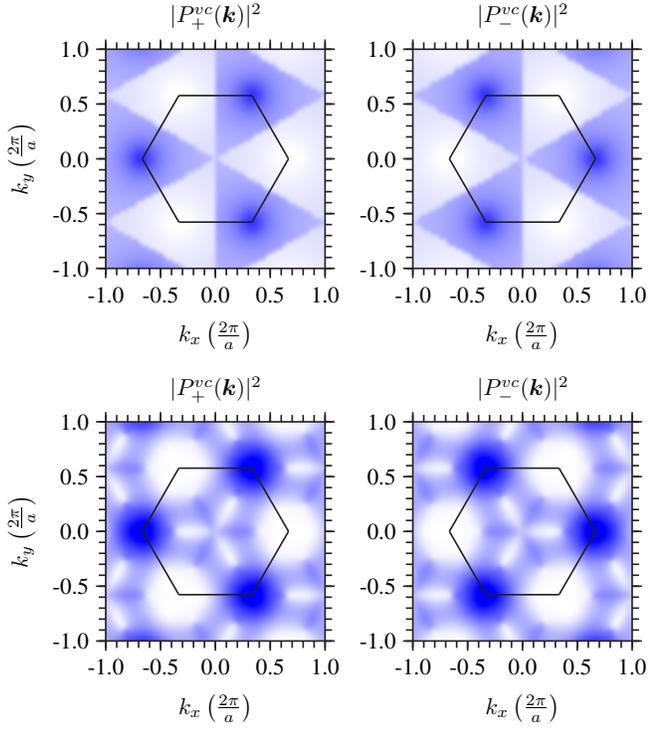}
\caption{Valence to lowest conduction band momentum matrix elements squared, for
circular polarization, for a linearized two-band model (top) and a non-linear
three-band model (bottom).  Black is positive and white is zero.  The results
are qualitatively very similar in the immediate vicinity of the $K$ and
$K^\prime$ points, but differ elsewhere in the Brillouin zone.
}
\label{fig:pcv_k_2}
\end{figure}

Valley-dependent selection rules have been shown to arise specifically for the
case of circularly polarized light~\cite{xia12}.  For circular polarizations,
the above expressions can be combined to give, in the two-band case,
\begin{equation}
\begin{split}
P^{vc}_{\pm}(\vk) &= \frac{1}{\sqrt{2}}\left[ P^{vc}_x(\vk) \pm i P^{vc}_y(\vk) \right] \\
    &= \mp \frac{mat}{\sqrt{2}\hbar} \left( 1 \mp \tau \frac{E_g}{2\varepsilon(\vk)} \right) 
        e^{\pm i\phi_\vk},
\end{split}
\end{equation}
leading to the valley-dependent intensities, 
\begin{equation}
\left| P^{vc}_{\pm}(\vk) \right|^2 = \frac{m^2 a^2 t^2}{2\hbar^2} \left( 1 \mp \tau \frac{E_g}{2\varepsilon(\vk)} \right)^2.
\end{equation}
Near the $K$ and $K^\prime$ points, $2\varepsilon(\vk) \rightarrow E_g$, such
that $P^{vc}_{\pm}(\vk) \propto (1\mp\tau)e^{\pm i \phi_\vk}$ and $\left|
P^{vc}_{\pm}(\vk) \right|^2 \propto (1\mp\tau)^2$, i.e.  circular polarization
can selectively excite electrons at the $K$ \textit{or} $K^\prime$ point. For
example, right-handed circular polarization, $P^{vc}_{-}(\vk)$, selectively
excites at the $K$ ($\tau = +1$) point.  Again, this analysis can be carried
out numerically for the three-band model.  A comparison of the the selection
rules, $\left|P^{vc}_{\pm}(\vk) \right|^2$, for the two models is shown in
Fig.~\ref{fig:pcv_k_2}.  Note that while the matrix elements themselves have an
ambiguity in the phase (i.e. they are not observable), the squared matrix
elements are completely independent of the phase convention.  In
Sec.~\ref{ssec:bse}, we will show how the nodal structure ($p$-type symmetry)
of the momentum matrix elements is canceled, leading to bright $s$-type
excitons \textit{which still respect the valley selectivity}.

\section{Linear optical properties and selection rules}
\label{sec:linear}

In general, the transition probability per unit time is given by
\begin{equation}\label{eq:transition}
W(\omega) = \frac{2\pi}{\hbar}\sum_{F}\left| V_{IF} \right|^2\delta(E_F-E_I-\hbar\omega)
\end{equation}
where $V_{IF}$ is the matrix element which couples the initial and final states
with energies $E_I$ and $E_F$. For the linear (one-photon) absorption, we have
$V = (eA/mc) \bm{\lambda} \cdot \hat{\vp}$, where $A$ is the vector potential
and $\bm{\lambda}$ is the polarization.  Within the presently considered model
Hamiltonians, symmetry excludes coupling to photons with electric vector
polarized perpendicular to the plane of the monolayer.  Here we explicitly
consider the case with electric vector polarized in the plane.  We will
evaluate this expression first in the independent particle picture and then in
the presence of excitonic effects.

\subsection{Independent particle absorption}
\label{ssec:ind_abs}

For an uncorrelated initial ground state $|0\rangle$ and an uncorrelated final
excited state $c^\dagger_{c,\vk} c_{v,\vk}|0\rangle$, it is simple to show
\begin{equation}
V_{IF} = \frac{eA}{mc} \langle 0 | \bm{\lambda} \cdot \hat{\vp} c^\dagger_{c,\vk} c_{v,\vk}|0\rangle
    = \frac{eA}{mc} \bm{\lambda} \cdot \vP^{vc}(\vk),
\end{equation}
\begin{equation}
E_F-E_I = E_c(\vk)-E_v(\vk),
\end{equation}
and therefore
\begin{equation}
\begin{split}
W(\omega) &= \frac{2\pi}{\hbar} \left(\frac{eA}{mc}\right)^2 \sum_{cv,\vk} 
        \left| \bm{\lambda}\cdot \vP^{vc}(\vk) \right|^2 \\
    &\hspace{8em} \times \delta(E_c(\vk)-E_v(\vk)-\hbar\omega).
\end{split}
\end{equation}
The imaginary part of the dielectric 
function~\footnote{Strictly, this is a 2D dielectric function, with units
of length, akin to a sheet polarizability per unit area.}
follows as~\cite{bas75}
\begin{equation}\label{eq:eps2}
\begin{split}
\varepsilon_2(\omega)  
    &= \frac{4\pi^2 e^2}{m^2 \omega^2} \sum_{cv} \int_{BZ}\frac{d^2k}{(2\pi)^2} 
        \left|\bm{\lambda}\cdot \vP^{vc}(\vk)\right|^2 \\
    &\hspace{8em} \times \delta(E_c(\vk)-E_v(\vk)-\hbar\omega),
\end{split}
\end{equation}
where we have taken the infinite-system limit.  Let us specifically consider
the linearized two-band model with right-handed circular polarization,
$\bm{\lambda}\cdot\vP^{vc}(\vk) = P^{vc}_-(\vk)$, for which we can carry out
the integration in Eq.~(\ref{eq:eps2}) semi-analytically.  Considering only one
valley (say $\tau = +1$), we can change to polar coordinates about the $K$
point,
\begin{equation}
\varepsilon_2^K(\omega) = \frac{2\pi e^2}{m^2 \omega^2} \int_0^\infty k |P^{vc}_-(k)|^2 \delta(2\varepsilon(k)-\hbar \omega) dk.
\end{equation}
Note that by integrating out to infinity, we are incurring an error at large
wavevectors (energies).  Since the dispersion relation is monotonic, we can
change variables, $k dk = \varepsilon d\varepsilon / a^2t^2$, and use the
squared matrix element from above to find 
\begin{equation}
\begin{split}
\varepsilon_2^K(\omega) &= \frac{\pi e^2}{\hbar^2 \omega^2} \int_0^\infty \hspace{-0.6em} d\varepsilon \hspace{0.1em} 
        \theta(2\varepsilon-E_g) \varepsilon \left(1 + \frac{E_g}{2\varepsilon}\right)^2 \delta(2\varepsilon-\hbar \omega) \\
    &= \frac{\pi e^2}{4\hbar \omega} \theta(\hbar\omega-E_g) \left(1 + \frac{E_g}{\hbar\omega}\right)^2. 
\end{split}
\end{equation}
Accounting for the other valley, $\varepsilon_2(\omega) =
\varepsilon_2^K(\omega) + \varepsilon_2^{K^\prime}(\omega)$, yields
\begin{equation}
\begin{split}
\varepsilon_2(\omega) 
    &= \frac{\pi e^2}{4\hbar \omega} \theta(\hbar\omega-E_g) \left[ \left(1 + \frac{E_g}{\hbar\omega}\right)^2 + \left(1 - \frac{E_g}{\hbar\omega}\right)^2 \right] \\
    &= \frac{\pi e^2}{2\hbar \omega} \theta(\hbar\omega-E_g) \left(1 +
    \frac{E_g^2}{(\hbar\omega)^2}\right).
\end{split}
\end{equation}
At energies just above the gap, the dielectric function is like that of a
conventional 2D semiconductor, i.e. $\omega^2 \varepsilon_2(\omega) = \mathrm{const}$,
but at higher energies it behaves like graphene (due to the linear dispersion),
i.e. $\omega \varepsilon_2(\omega) = \mathrm{const}$.  However, the linear dispersion is
unrealistic for TMDCs, as can be seen in the full band structure
(Fig.~\ref{fig:bands}).

\subsection{Exciton absorption and the Bethe-Salpeter equation}
\label{ssec:bse}

We now consider the spin-singlet optical properties including the excitonic
effects arising from the strong electron-hole interaction.  The
\textit{correlated} excited states within the single-excitation approximation
can be written as
\begin{equation}\label{eq:wavefunction}
|X\rangle = \sum_{\vk}^{BZ} \sum_{vc} 
    A_{vc}^X(\vk)\ c^\dagger_{c,\vk} c_{v,\vk} |0\rangle, 
\end{equation}
where $|0\rangle$ is again an uncorrelated (determinental) ground state.  This
form for the excited state wavefunction underlies the time-dependent
Hartree-Fock and Bethe-Salpeter equation (BSE) formalisms; here we will pursue
the latter, which is a many-body perturbative theory in the screened
two-particle interaction.  For a periodic crystal exciton wavefunction,
Eq.~(\ref{eq:wavefunction}), the BSE is an eigenvalue problem~\cite{roh00} for
the exciton energy $E_X$, 
\begin{widetext}
\begin{equation}\label{eq:bse}
E_X A_{vc}^X(\vk) = 
\left( E_{c,\vk}-E_{v,\vk} \right) A_{vc}^X(\vk) 
    + \frac{1}{A} \sum_{\vk^\prime}^{BZ} \sum_{v^\prime,c^\prime} 
        \langle \psi_{v,\vk} \psi_{c,\vk} | K^{eh} | 
            \psi_{v^\prime,\vk^\prime} \psi_{c^\prime,\vk^\prime} \rangle 
        A_{v^\prime c^\prime}^X(\vk^\prime).
\end{equation}
The electron-hole interaction kernel $K^{eh}$ is the sum of a
frequency-dependent screened Coulomb interaction and an unscreened exchange
interaction~\cite{roh00},
\begin{subequations}
\begin{align}
\langle \psi_{v,\vk} \psi_{c,\vk} | K^{eh,d} | 
        \psi_{v^\prime,\vk^\prime} \psi_{c^\prime,\vk^\prime} \rangle
    &= - \int d^d\vr \int d^d \vr^\prime \psi_{c,\vk}^*(\vr) \psi_{c^\prime,\vk^\prime}(\vr) 
        W(\vr,\vr^\prime,\omega) 
            \psi_{v,\vk}(\vr^\prime) \psi_{v^\prime,\vk^\prime}^*(\vr^\prime) \\
\langle \psi_{v,\vk} \psi_{c,\vk} | K^{eh,x} | 
        \psi_{v^\prime,\vk^\prime} \psi_{c^\prime,\vk^\prime} \rangle
    &= \int d^d\vr \int d^d \vr^\prime \psi_{c,\vk}^*(\vr) \psi_{v,\vk}(\vr)
    |\vr-\vr^\prime|^{-1}
            \psi_{c^\prime,\vk^\prime}(\vr^\prime)\psi_{v^\prime,\vk^\prime}^*(\vr^\prime). 
\end{align}
\end{subequations}

If we (i) neglect the exchange interaction, (ii) neglect the
frequency-dependence and local-field effects of the screened direct
interaction, i.e. $W(\vr,\vr^\prime,\omega) \approx
W(\vr-\vr^\prime,\omega=0)$, and (iii) make a `zero differential overlap'
approximation for the atomic orbitals, we find
\begin{equation}\label{eq:phase_screened}
\langle \psi_{v,\vk} \psi_{c,\vk} | K^{eh} | 
        \psi_{v^\prime,\vk^\prime} \psi_{c^\prime,\vk^\prime} \rangle
    \approx - \langle \psi_{c,\vk} | \psi_{c^\prime,\vk^\prime} \rangle 
        \langle \psi_{v^\prime,\vk^\prime} | \psi_{v,\vk} \rangle W(\vk-\vk^\prime).
\end{equation}
In the above, we have neglected the possible orbital structure to the screened
interaction $W_{ij}(\vr-\vr^\prime)$.

\end{widetext}

At this point, we wish to emphasize that the orbital overlap prefactor in the
screened interaction is crucially important. As an explicit example, in the
two-band picture, we have
\begin{subequations}
\begin{align}
\begin{split}
\langle\psi^\tau_{c,\vk}|\psi^\tau_{c,\vk^\prime}\rangle &= \frac{1}{2}\Big[
\sqrt{\alpha_+(\vk)\alpha_+(\vk^\prime)} \\
    &\hspace{2em} + \sqrt{\alpha_-(\vk) \alpha_-(\vk^\prime)}
    e^{-i\tau(\phi_\vk-\phi_{\vk^\prime})} \Big],
\end{split}\\
\begin{split}
\langle\psi^\tau_{v,\vk^\prime}|\psi^\tau_{v,\vk}\rangle &= \frac{1}{2}\Big[
\sqrt{\alpha_-(\vk^\prime)\alpha_-(\vk)} \\
    &\hspace{2em} + \sqrt{\alpha_+(\vk^\prime) \alpha_+(\vk)}
    e^{i\tau(\phi_\vk-\phi_{\vk^\prime})} \Big]. 
\end{split}
\end{align}
\end{subequations}
As before, near the $K$ and $K^\prime$ points,  $2\varepsilon(\vk) \rightarrow
E_g$, [i.e. $\alpha_+(\vk) \approx 1$ and $\alpha_-(\vk) \approx 0]$, and in
this limit, 
\begin{subequations}\label{eq:overlaps}
\begin{align}
\langle\psi^\tau_{c,\vk}|\psi^\tau_{c,\vk^\prime}\rangle &\approx 1 \\
\langle\psi^\tau_{v,\vk^\prime}|\psi^\tau_{v,\vk}\rangle &\approx e^{i\tau(\phi_\vk-\phi_{\vk^\prime})}. 
\end{align}
\end{subequations}
The BSE, Eq.~(\ref{eq:bse}), then yields a Wannier-\textit{like}, two-band
picture with an unusual phase factor in the screened interaction,
\begin{equation}\label{eq:bse_phase}
\begin{split}
E_X A_{vc}^X(\vk) &= \left( E_{c,\vk}-E_{v,\vk} \right) A_{vc}^X(\vk) \\
    &\hspace{1em} - \frac{1}{A}\sum_{\vk^\prime}^{BZ} W(\vk-\vk^\prime) 
        e^{i\tau(\phi_\vk-\phi_{\vk^\prime})} A_{vc}^X(\vk^\prime).
\end{split}
\end{equation}
Multiplying through by $e^{-i\tau\phi_\vk}$ gives a conventional Wannier
equation for the pseudo-wavefunction $\tilde{A}_{vc}^X(\vk) = e^{-i\tau
\phi_\vk} A_{vc}^X(\vk)$.  If the bands can be approximated as parabolic, this
means that the energy \textit{spectrum} of the BSE is identical to that of a
corresponding real-space Wannier equation with a screened interaction $W(\vr)$,
as we have employed in previous work~\cite{ber13,che14},
\begin{equation}\label{eq:wannier}
\left[E_X - E_g\right] \tilde{A}_{vc}^X(\vr).
    = \left[-\frac{1}{2\mu} \nabla_\vr^2 - W(r)\right] \tilde{A}_{vc}^X(\vr) 
\end{equation}
However, as explained in a recent work by Srivastava and 
Imamoglu~\cite{sri15}, systematically continuing the expansion of 
Eqs.~(\ref{eq:overlaps}) for small $k-k^\prime$ leads to additional terms in the 
Coulomb interaction that weakly break certain degeneracies (see below). In this
case, the spectrum of Eqs.~(\ref{eq:bse_phase}) and (\ref{eq:wannier}) is no
longer identical to that of the BSE with the screened interaction given in
Eq.~(\ref{eq:phase_screened}).

It remains to be shown whether the exciton wavefunctions of the original
problem, as described by the BSE (\ref{eq:bse_phase}), have the same selection
rules or the same spatial symmetries as the wavefunction of the real-space
Wannier equation (\ref{eq:wannier}).  To analyze the spatial symmetries, we can
calculate the real-space wavefunction corresponding to the solution of the BSE,
with the hole position fixed at the origin.  We find
\begin{equation}
\begin{split}
\Psi_X(\vr_e,\vr_h=0) &\equiv \sum_\vk A_{vc}^X(\vk) \psi_{c,\vk}(\vr_e) \psi^*_{v,\vk}(0) \\
    &\approx \sum_\vk A_{vc}^X(\vk) e^{-i\tau\phi_\vk}  e^{i\vk\cdot\vr_e} 
    = \tilde{A}_{vc}^X(\vr_e),
\end{split}
\end{equation}
demonstrating that the wavefunction which solves the real-space
Eq.~(\ref{eq:wannier}) is indeed (approximately) the same as the real-space BSE
wavefunction. At a less approximate level, the spatial \textit{symmetries}
($s$, $p$, $d$, etc.) will be identical.  This is one of the main conclusions
of this work.  

To determine the selection rules, we now consider the optical absorption in the
presence of correlated excitonic effects.  Assuming as before an uncorrelated
initial (ground) state $|I\rangle = |0\rangle$, but now using a Wannier-like
final exciton state $|X\rangle$ as in Eq.~(\ref{eq:wavefunction}) gives
\begin{equation}
\langle I|\bm{\lambda} \cdot \hat{\vp}|X\rangle = \sum_\vk A^{X}_{vc}(\vk) 
\bm{\lambda}\cdot \vP^{vc}(\vk),
\end{equation}
which leads to the dielectric function
\begin{equation}\label{eq:eps2_bse}
\varepsilon_2(\omega) = \frac{4\pi^2e^2}{m^2\omega^2} \sum_X \left| \sum_\vk A_{vc}^X(\vk)
\bm{\lambda}\cdot \vP^{vc}(\vk) \right|^2 \delta(\hbar\omega-E_X).
\end{equation}
Recall that for right-handed circular polarization, the momentum matrix element
near the $K^\prime$ ($\tau=-1$) point is nearly zero and near the $K$
($\tau=1$) point it is given by 
\begin{equation}
\bm{\lambda}\cdot\vP^{vc}(\vk) = P^{vc}_-(\vk) \approx \frac{\sqrt{2} mat}{\hbar} e^{-i\phi_\vk}\equiv P_0 e^{-i\phi_\vk}.
\end{equation}
Therefore we can restrict our attention to $\vk$ near $K$, which gives 
\begin{equation}\label{eq:Vif_X}
\langle I|\bm{\lambda} \cdot \hat{\vp}|X\rangle = P_0\sum_{\vk\sim K} A^{X}_{vc}(\vk) e^{-i\phi_\vk} = P_0 \tilde{A}^{X}_{vc}(\vr=0) 
\end{equation}
and therefore
\begin{equation}
\varepsilon_2(\omega) = \frac{4\pi^2e^2P_0^2}{m^2 \omega^2} \sum_X \left| \tilde{A}_{vc}(\vr=0) \right|^2 \delta(\hbar\omega-E_X),
\end{equation}
which is just the usual Elliott formula for the excitonic
absorption~\cite{ell57}.  In particular, the selection rules are conventional
in that they are determined by the behavior of the wavefunction at the origin
in real-space, leading to bright states with $s$-type azimuthal symmetry.  We
emphasize that the phase factor appearing in the momentum matrix element is
essentially cancelled by the conjugate phase factor in the exciton envelope
wavefunction, which itself originates from the change of basis in the screened
interaction, Eq.~(\ref{eq:phase_screened}).  Therefore, not only can the
excitons be labeled in analogy with the hydrogen series in terms of their
spatial symmetries but, to lowest order, they also obey identical selection
rules.  This is the second main conclusion of this work.

\begin{figure}[t!]
\centering
\includegraphics[scale=1.0]{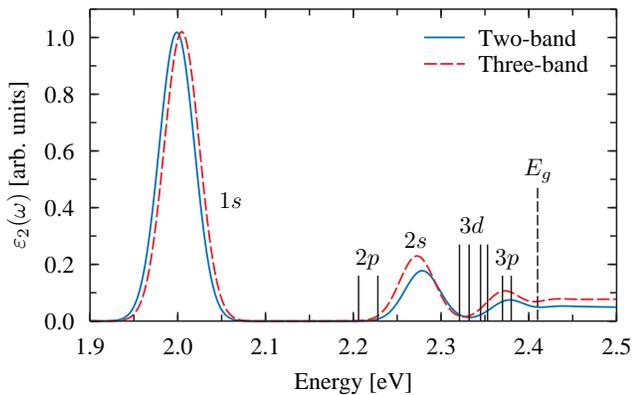}
\caption{
Imaginary part of the dielectric function for MoS$_2$ calculated in the
presence of excitonic effects.  The band gap has been rigidly increased to 2.41
eV such that the $1s$ exciton peak occurs near 2.0 eV (spin-orbit splitting
into $A$ and $B$ peaks is neglected, as described in the text). A Gaussian
broadening of 50 meV (FWHM) has been applied to all peaks. 
}
\label{fig:eps_bse}
\end{figure}

As usual, the same analysis cannot be done analytically on the three-band
model, but it can be straightforwardly carried out numerically.  The dielectric
function calculated via Eq.~(\ref{eq:eps2_bse}) for the two considered band
structure models of MoS$_2$ is plotted in Fig.~\ref{fig:eps_bse}; in
particular, the orbital overlaps in Eq.~(\ref{eq:phase_screened}) are
calculated numerically, without the approximation given in
Eqs.~(\ref{eq:overlaps}). As described in Refs.~\cite{ber13,che14}, the
screened interaction used in the calculations is given by
\begin{equation}\label{eq:screened_int}
W(\vk) = \frac{2\pi e^2}{k(1 + 2\pi \chi_{2D} k)}
\end{equation}
with $\chi_{2D} = 6.6$ \AA\ for intrinsic MoS$_2$.  Results are presented for a
$120\times120$ sampling of the Brillouin zone, which we have found necessary to
converge the binding energy to roughly 0.1 eV accuracy, in agreement with the
fully \textit{ab initio} BSE study presented in Ref.~\cite{qiu13}.
Specifically, for MoS$_2$ this sampling gives a $1s$ exciton binding energy of
0.41 eV, however an extrapolation to the infinite sampling limit gives
approximately 0.52 eV, in good agreement with our prior result obtained in
Ref.~\cite{ber13} (0.54 eV).  In Fig.~\ref{fig:eps_bse}, the conduction
bands have been rigidly shifted to increase the band gap to 2.41 eV, such that
the $1s$ exciton peak occurs near its experimentally observed value of 2.0 eV
(due to the spin-orbit interaction, this peak is actually split into the
so-called $A$ and $B$ peaks at about 1.9 and 2.0 eV respectively~\cite{mak10}).
An important conclusion to be drawn from Fig.~\ref{fig:eps_bse} is that the
more realistic band structure generates only minor quantitative differences in
$\varepsilon_2(\omega)$, compared to that generated by the two band model.

\begin{figure}
\centering
\includegraphics[scale=0.35]{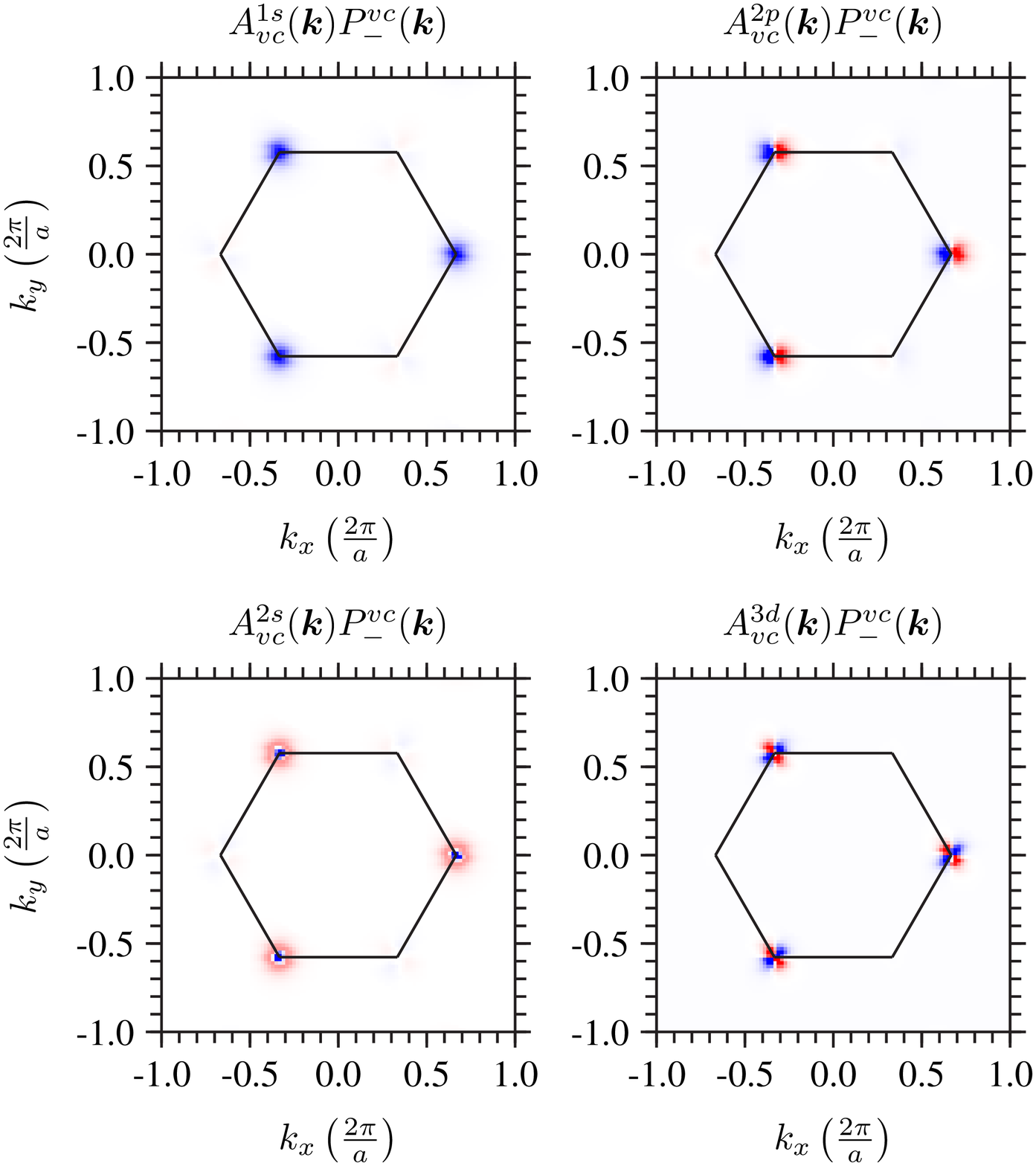}
\caption{Reciprocal space plots of the selection-rule-determining product
$A^X_{vc}(\vk)P_-^{vc}(\vk)$. In the presence of right-handed circular
polarization, it is seen that excitons are only created at the $K$ point, and
not at the $K^\prime$ point, as was found in Ref.~\cite{xia12} and in
Sec.~\ref{ssec:trans} in the \textit{absence} of exciton effects.
}
\label{fig:wfns}
\end{figure}

The labeling of states in Fig.~\ref{fig:eps_bse} is done via inspection of the
wavefunction, in either reciprocal or real-space.  For example, in
Fig.~\ref{fig:wfns} we show the selection-rule-determining product
$A^X_{vc}(\vk)P_-^{vc}(\vk)$ [which is closely related to the pseudo-wavefunction
$\tilde{A}^X_{vc}(\vk)$] for right-handed polarization.  The symmetries of the
exciton wavefunctions are apparent, and the valley selectivity is also
recovered in the presence of excitonic effects.  

Focusing on the features in the $\varepsilon_2(\omega)$ spectrum that derive
from the $s$-type exciton states, the Rydberg series is nonhydrogenic, as
discussed in detail in Refs.~\cite{che14,hil15}.  This follows from the
unusual form of the screened Coulomb interaction for these monolayer thick
materials.  In particular, it deviates substantially from the $1/\varepsilon_0
r$ form that dominates in conventional semiconductors.  The Hamiltonian with
this latter interaction has additional symmetry which leads to the
``accidental'' angular momentum degeneracy in the hydrogen spectrum.  Here that
symmetry is broken: we find that for a given principal quantum number, the
larger angular momentum states are more strongly bound, i.e.  $E_{1s} < E_{2p}
< E_{2s} < E_{3d}$ and so on.  The same behavior has been recently observed in
a fully \textit{ab initio} BSE calculation~\cite{ye14}, and the present work
provides a simple physical explanation for this behavior in terms of the
effective screened interaction (see also Refs.~\cite{wu15,cha15} for
similar findings).  To verify this unconventional disposition of dark exciton
states requires a nonlinear spectroscopic measurement, which we discuss in the
next section.  Furthermore, we also note a small splitting of the $2p$, $3d$, and
$3p$ dark exciton states. In particular, the 20 meV splitting of the $2p$
states is in good agreement with recent results~\cite{wu15,sri15}. As mentioned
before, Srivastava and Imamoglu have traced this degeneracy breaking to the
orbital overlaps in Eq.~(\ref{eq:phase_screened}) and explained the effect in terms
of Berry curvature in the single-particle band structure~\cite{sri15}.

\section{Two-photon absorption}
\label{sec:tpa}

Our theoretical framework for the two-photon absorption essentially follows the
early work of Mahan~\cite{mah68} for 3D semiconductors and Shimizu~\cite{shi89}
for 2D quantum wells including explicit consideration of excitons.  For a
two-photon process, the transition rate is again given by
Eq.~(\ref{eq:transition}), except we now have two perturbing fields, $V_i = (eA_i/mc)
\bm{\lambda}_i \cdot \hat{\vp}$ ($i=1,2$), where $A_i$ is the vector potential,
$\bm{\lambda}_i$ is the polarization, and $\hbar \omega_i$ is the photon
energy. The matrix element of the perturbation can be evaluated by a sum over
intermediate states $|M\rangle$,
\begin{equation}\label{eq:tpa}
\begin{split}
V_{IF} &= \left(\frac{e}{mc}\right)^2 A_1 A_2 \sum_M 
 \Bigg[\frac{\langle I|\bm{\lambda}_2 \cdot \hat{\vp}|M\rangle\langle M|\bm{\lambda}_1 
        \cdot \hat{\vp}|F\rangle} {E_M-E_I-\hbar \omega_1} \\
&\hspace{8em}+ \frac{\langle I|\bm{\lambda}_1 \cdot \hat{\vp}|M\rangle\langle
M|\bm{\lambda}_2 
        \cdot \vp|F\rangle}{E_M-E_I-\hbar \omega_2} \Bigg].
\end{split}
\end{equation}
The two-photon spectroscopy of single-particle states is trivial, and so we
restrict our analysis to the excitonic case.  As in the one-photon exciton
absorption, Eq.~(\ref{eq:Vif_X}) holds for the matrix element connecting the
ground and \textit{intermediate} exciton states.  In contrast, the matrix
element between two exciton states (intermediate and final) is
\begin{equation}\label{eq:int_final}
\begin{split}
\langle M|\bm{\lambda}_1 \cdot \hat{\vp}|F\rangle &= \hbar\sum_{\vk} A^{M*}_{vc}(\vk) 
        \bm{\lambda}_1\cdot \vk A^F_{vc}(\vk) \\
    &= \hbar\sum_{\vk} \tilde{A}^{M*}_{vc}(\vk)e^{-i\tau\phi_\vk}
        \bm{\lambda}_1\cdot \vk \tilde{A}^F_{vc}(\vk)e^{i\tau\phi_\vk}\\
    &= -i\hbar \int d^2r \tilde{A}^{M*}_{vc}(\vr) \bm{\lambda}_1\cdot\nabla_{\vr}\tilde{A}^F_{vc}(\vr).
\end{split}
\end{equation}
In the above, we have restricted the analysis to two bands ($c,v$) and used the
facts that the expectation value of $\hat{\vp}$ is zero in a Slater determinant
and that $\hat{\vp}$ is diagonal in reciprocal space.  To have a nonzero
Eq.~(\ref{eq:int_final}) requires that the real-space exciton wavefunctions
$A^F$ and $A^M$ have orbital angular momenta which differ by $\pm 1$; this is
the same two-photon selection rule as found in conventional semiconductors
including consideration of exciton effects.  Combined with the result of the
previous section -- that one-photon absorption produces $s$-type excitons -- we
conclude that two-photon absorption produces only $p$-type excitons.  With
these results, the two-photon absorption essentially follows the early work of
Mahan~\cite{mah68} for 3D semiconductors or Shimizu~\cite{shi89} for 2D quantum
wells.  

The primary complication in the evaluation of two-photon absorption is the
evaluation of the internal sum over intermediate states in Eq.~(\ref{eq:tpa}).
We follow the approximation introduced by Mahan~\cite{mah68} and used by
Shimizu~\cite{shi89} that allows the sum to be eliminated with a completeness
relation.  Explicitly incorporating the above results, the first term in
Eq.~(\ref{eq:tpa}) can be written as  (the second term is analogous)
\begin{equation}
\begin{split}
-i\hbar P_0 &\int d^2r \sum_M\frac{\tilde{A}^M_{vc}(\vr=0)
\tilde{A}^{M*}_{vc}(\vr)}{E_M-E_I-\hbar\omega_1} \bm{\lambda}_1\cdot \nabla_\vr
\tilde{A}^F_{vc}(\vr) \\
    &\approx \frac{-i\hbar P_0}{E_g-\langle E_b\rangle-\hbar\omega_1}
    \left[\bm{\lambda}_1 \cdot \nabla_\vr \tilde{A}^F_{vc}(\vr)\right]_{\vr=0}
\end{split}
\end{equation}
where $\langle E_b\rangle$ is an average intermediate ($s$-type) exciton energy
introduced to facilitate the (complete) sum over intermediate states; for
simplicity we will henceforth set $\langle E_b \rangle$ to zero as its primary
influence is to simply alter the prefactor.  In contrast to the hydrogenic
exciton case, where further results can be obtained analytically, the matrix
elements here must be evaluated numerically.

The two-photon transition rate is thus given by
\begin{equation}
\begin{split}
W(\Omega) &= 2\pi\hbar\left(\frac{e}{mc}\right)^4 \left(A_1 A_2\right)^2 (\hbar P_0)^2 \\
    &\times \sum_F \left| \frac{\left[ \bm{\lambda}_1 \cdot \nabla_\vr 
                    \tilde{A}^F_{vc}(\vr)\right]_{\vr=0}}
                {E_g -\hbar \omega_1} +\{1\leftrightarrow 2\} \right|^2
    \delta(\hbar\Omega - E_F)
\end{split}
\end{equation}
where $\hbar\Omega = \hbar\omega_1 + \hbar\omega_2$.  The simplest case to
consider is when $\bm{\lambda}_1 = \bm{\lambda}_2$ and
$\hbar\omega_1 = \hbar\omega_2 \approx E_g/2$, which gives
\begin{equation}\label{eq:tpa_final}
W(\Omega) = W_0 \sum_F \left| \bm{\lambda} \cdot \nabla_\vr 
        \tilde{A}^F_{vc}(\vr) \right|^2_{\vr=0}
    \delta(\hbar\Omega - E_F)
\end{equation}
where
\begin{equation}
W_0 = \frac{32 \pi \hbar^3 e^4 A_1^2 A_2^2 P_0^2}{m^4 c^4 E_g^2}.
\end{equation}
If both photons have the same circular polarization, then this experiment
probes valley-selective $p$-type excitons, which are dark in the linear
measurement.  Using photons with opposite polarizations would create $p$-type
excitons in both valleys.

\begin{figure}
\centering
\includegraphics[scale=1.0]{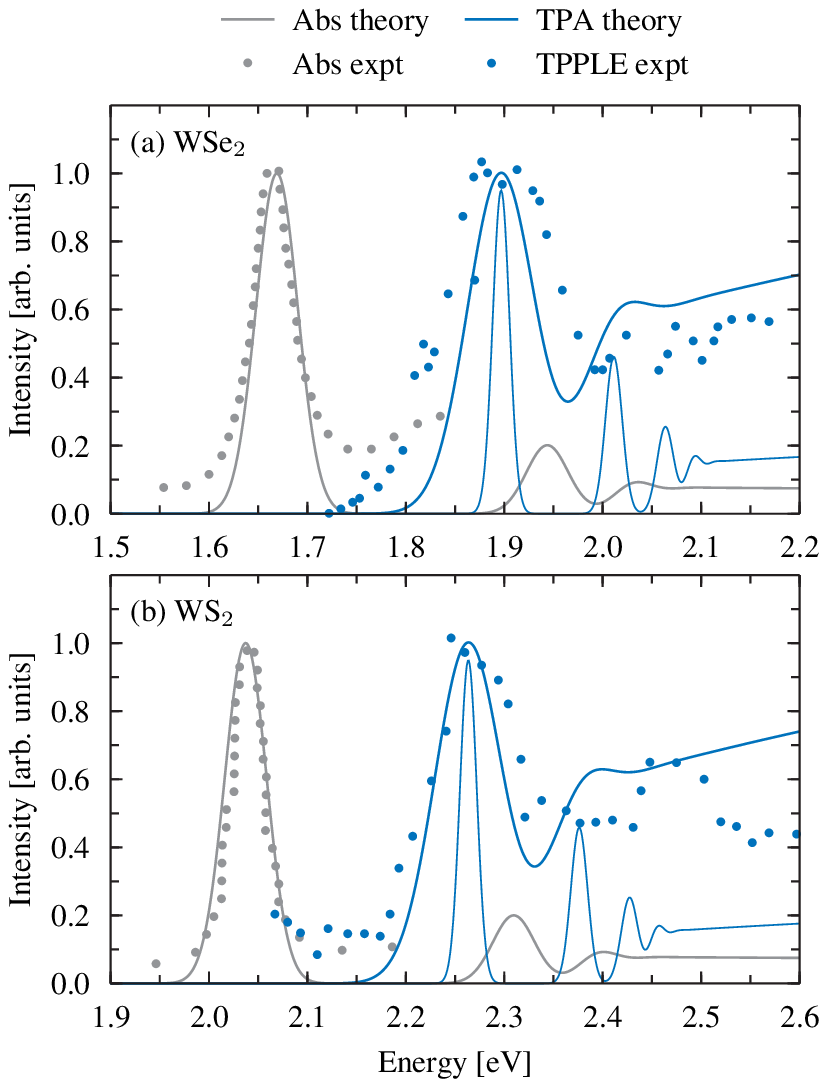}
\caption{
Two-photon absorption (TPA) intensity for monolayer (a) WSe$_2$ and (b) WS$_2$
evaluated numerically with Eq.~(\ref{eq:tpa_final}) (blue lines).  The spectra
have been artificially broadened with a Gaussian linewidth (FWHM) of 80 meV
(thicker line) and 20 meV (thinner line).  The experimental two-photon
photoluminescence excitation (TPPLE) spectrum for WSe$_2$~\cite{he14} and
WS$_2$~\cite{ye14} is included for comparison (blue circles).  The theoretical
linear absorption spectrum from the same model (FWHM of 50 meV) is overlaid for
reference (grey lines) along with the experimental result (gray circles) for
WSe$_2$~\cite{he14} and WS$_2$~\cite{li14}.
}
\label{fig:tpa}
\end{figure}

Motivated by recent nonlinear spectroscopic measurements on WSe$_2$~\cite{he14}
and WS$_2$~\cite{ye14}, in Fig.~\ref{fig:tpa} we show the results of a
numerical evaluation of Eq.~(\ref{eq:tpa_final}) for these two materials; the
exciton wavefunctions and their derivatives have been obtained from the
real-space effective mass treatment of the two-band model (i.e.~the small
splitting of the $p$ excitons is neglected).  The agreement with experiment,
for both the linear and nonlinear response, is seen to be quite good.  In the
calculations, we have used the same screening length, $\chi_{2D} = 7.0$~\AA\
for both materials, which yields an exciton binding energy of 0.48 eV (in
accord with our previous results~\cite{ber13}).  We note that this exciton
binding energy is slightly larger than that determined in
Refs.~\cite{he14,che14} (0.37 and 0.32 eV for WSe$_2$ and WS$_2$
respectively).

In the narrow linewidth limit, the two-photon absorption identifies the
$p$-type excitons with energies slightly below that of the corresponding
$s$-type exciton.  For a larger linewidth, the $2p$ transition is still
resolved and responsible for the main peak seen in experiment, while the
remaining transitions merge to yield a weak feature before the continuum onset.
Importantly, ratio between the $2p$ peak height and the higher-energy signal
(near the continuum onset) is determined by the spectral linewidth.  It is thus
encouraging that our simulated spectrum simultaneously reproduces the $2p$
linewidth and this intensity ratio; the required broadening suggests that it
should be difficult to observe the $3p$ transition at this resolution.  This
leaves open the origin of the small feature observed near 2.5~eV in the
experimental spectrum for WS$_2$.

Finally, we point out that a recent study on WSe$_2$ using one- and two-photon
photoluminescence excitation spectroscopy~\cite{wan15}, has identified the 2$s$
and 2$p$ transitions to have the same energy to meV accuracy.  This is in quite
stark contrast with the results of the present work, which suggest that the
2$p$ exciton should be lower in energy by \textit{at least} 50 meV.  We hope
that future work, both experimental and theoretical, is devoted to
investigating this discrepancy.


\section{Conclusions}
\label{sec:conc}

In this work, we have expanded the effective mass theory presented in
Refs.~\cite{ber13,che14} to include a fully $k$-dependent model of the
band structure, in harmony with other recent works~\cite{berg14,kon14,wu15}.
This extension allows for deviations from parabolicity, including trigonal
warping behavior which has been emphasized in other
contexts~\cite{kor13,rost13}.  We find that two- and three-band models of the
single-particle band structure give nearly identical results for the exciton
properties within a simplified BSE formalism, suggesting that trigonal warping
is a secondary effect.  Furthermore, our numerical results are nearly identical
to those of the effective mass treatment from our previous
work~\cite{ber13,che14}, justifying its use in those contexts.  We have
definitively proved that spin-singlet excitons with $s$-type azimuthal
symmetry, which have been the most studied~\cite{ber13,berg14,che14}, are indeed
the optically bright excitons.  As in our previous work~\cite{che14,hil15}, we
confirm that the disposition of bright exciton states is distinctly
non-hydrogenic.

The dark spin-singlet excitons have also been investigated and found to exhibit
another deviation from the hydrogen model, in the form of a broken angular
momentum degeneracy. Using an approach similar to ours, the authors of
Refs.~\cite{wu15,cha15} have identified the same qualitative behavior.
This observation will be key in future analyses of two-photon spectroscopies on
TMDCs.  A recent manuscript contains results from a fully \textit{ab initio}
BSE calculation on WS$_2$ and also finds this peculiar angular momentum
behavior~\cite{ye14}.  It is clearly encouraging that our simple low-energy
theory -- featuring a few-band representation of the single-particle states and
an appropriate treatment of screening with a model dielectric function -- is
able to correctly reproduce the optical selection rules, the character of
bright and dark exciton states, the broken angular momentum degeneracy, the
quantitatively large exciton binding energies, and the spectral features of the
nonlinear two-photon absorption.  In this regard, we believe the model
presented here represents perhaps the simplest predictive minimal model capable
of unifying these wide-ranging features in monolayer TMDCs.

\textit{Note added.} As discussed in the main text, a recent preprint
analyzes the impact of the band overlap factors in the effective Coulomb
interaction, Eq.~(\ref{eq:phase_screened}), and systematically develops the
next order terms in $k-k^\prime$, demonstrating signatures of the Berry
curvature in the exciton spectra~\cite{sri15}.  Our numerical results agree
with their analysis and with their estimate for the splitting of the $2p$
exciton levels. Figure~\ref{fig:eps_bse} was updated to reflect these splittings.

\begin{acknowledgments} 
The authors would like to thank Alexey Chernikov and Tony F. Heinz for
invaluable discussions. T.C.B~thanks Ajit Srivastava for informative correspondence
regarding Ref.~\cite{sri15}. Part of this work was supported by the Princeton Center
for Theoretical Science (TCB), and part of this work was done using resources of 
the Center for Functional Nanomaterials, which is a U.S. DOE Office of
Science User Facility, at Brookhaven National Laboratory under Contract No.
DE-SC0012704 (MSH).  
\end{acknowledgments}

\end{document}